# From Citation Intent to Knowledge Contribution: Classifying What Cited Papers Actually Contribute

Zhibang Quan, Zhentao Liang, Ming Ma, Jinyu Wei, Gang Li, Jin Mao*
Center for Studies of Information Resources, Wuhan University, Wuhan, 430072, China
School of Information Management, Wuhan University, Wuhan, 430072, China

**Abstract:**

Understanding the flow and evolution of scientific knowledge is essential for assessing research impact. Existing citation analysis methods mainly focus on citing authors' subjective intents, failing to consistently characterize cited papers' knowledge contributions. This study proposes the Knowledge Contribution Taxonomy (KCT), derived from the Scientific Research Logic Model, which identifies the type of knowledge a cited paper contributes based on the citation context. KCT classifies citations into Method, Resource Tool, Empirical Finding, and Background, further distinguishing core from non-core contributions. We propose a Dual-Path Fusion model for the classification task, which achieves an accuracy of 85.5%, outperforming mainstream large language models. An analysis of 802,202 citations from the ACL Anthology reveals that core knowledge contributions account for only 39.09% of all citations. The core knowledge contribution citation count achieves higher hit rates for award-winning papers than the traditional citation count at all ranking cutoffs, reflecting the value of differentiating knowledge contributions for research evaluation and impact prediction. In dissemination prediction experiments, KCT outperforms citation intent classification, demonstrating its stronger predictive validity for scholarly dissemination. By focusing on the knowledge contributions of cited papers, the KCT can support differentiated research evaluation.

* Corresponding Author. Email: danveno@163.com

# 1. Introduction

Understanding how knowledge propagates and accumulates in literature is a central concern of bibliometrics and science of science (Aman & Gläser, 2025). Citations serve as the primary vehicle for this process, reflecting the way that researchers incorporate methods, data, findings, and theories from prior work into their own. Since Garfield (1955) proposed citation indexing, citation counts have become the dominant measure of scholarly impact, institutionalized through tools such as Impact Factor and h-index, which are based on the implicit assumption that every citation constitutes an equivalent signal of influence. In practice, however, citation behavior is inherently heterogeneous. Studies have shown that citation motivations are diverse and that not all citations represent equal degrees of knowledge input (Bornmann & Daniel, 2008; Tahamtan & Bornmann, 2019). Many citations are rhetorical or perfunctory rather than reflecting substantive engagement with the cited work (Bornmann & Leibel, 2026). This heterogeneity has driven a shift in citation research from counting citations to demystifying the internal mechanisms of citation behavior.

Citation intent classification has been the primary approach to understanding citation behavior, aiming to identify the functional role of a citation by analyzing its citing context (Cohan et al., 2019; Jurgens et al., 2018), with continued improvements in classification performance in recent years (Fogelson et al., 2025; Kunnath et al., 2023; Lauscher et al., 2022). However, there are fundamental challenges in intent classification. Intent judgments rely on inferring the citing author's purpose, yet the complexity of citation behavior renders such inference inherently uncertain (Erikson & Erlandson, 2014), and different annotators frequently disagree on the same citation (Kunnath et al., 2021). More critically, citation intent captures a citation's function within a specific passage, making intent classification frameworks inherently unable to characterize what knowledge the cited work contributes. Although Duan and Tan (2026) noted this conflation between citation intent and cited content type, their work remains centered on the citing context and does not analyze the cited paper's knowledge contributions directly.

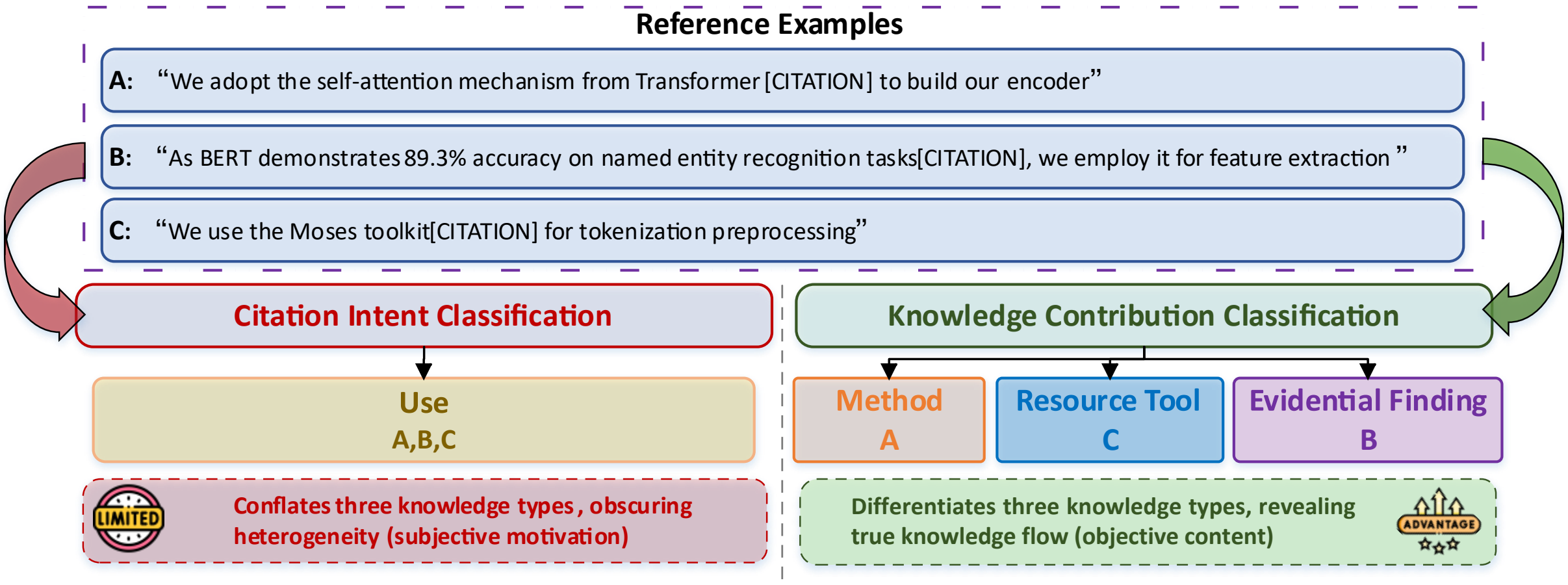

**Figure 1.** Comparing Intent-Based and Contribution-Based Citation Classification

Unlike citation intents, the concept of knowledge contribution refers to what type of knowledge the cited paper contributes to the citing paper. Figure 1 provides a concrete example.

The three citations adopt the self-attention mechanism, entity recognition accuracy, and tokenization tool from the cited paper, respectively. When classified by intents, all three are labeled as “Use,” conflating three different types of knowledge input. From the knowledge contribution perspective, however, each citation is classified according to the specific type of knowledge it transmits. This distinction carries significant analytical implications, as different types of knowledge may experience different dissemination pathways across the scientific community (Aman & Gläser, 2025). Yet existing citation classification research primarily infers citing motivations from the citing author's perspective, leaving the knowledge contributions of cited works largely unexamined.

Recent studies have begun to explore the typological identification of scientific contributions. Chen et al. (2025) classified citation contexts to profile the contributions of Nature and Science papers and examined their relationships with the division of labor among authors, finding a notable discrepancy between actual contributions as perceived by peers and those self-identified by authors. Pramanick et al. (2025) proposed an eight-category taxonomy to extract and track author-stated contributions across five decades of NLP research. However, these studies remain limited to contribution identification and distribution description, and do not further examine how different knowledge types differ in dissemination behavior and what this implies for research evaluation. It is therefore necessary to construct a classification framework that can identify the knowledge contribution types of references, and to leverage it for impact assessment and dissemination prediction, as well as to validate the effectiveness of the knowledge contribution perspective in research evaluation.

To this end, we first conceptualize the Scientific Research Logic Model. Based on this model, we propose the Knowledge Contribution Taxonomy (KCT), which identifies the type of knowledge that a cited paper contributes to the citing paper. We develop an automatic classification model and apply it to 802,202 citations in the ACL Anthology (1980–2024) to analyze the structural characteristics of knowledge flow. Building on this, we apply the KCT to impact assessment and dissemination prediction, and compare its performance against intent classification, validating the effectiveness of the knowledge contribution perspective in research evaluation. Our contributions are summarized as follows:

1) We propose the Knowledge Contribution Taxonomy (KCT), which comprises four categories: Method, Resource Tool, Empirical Finding, and Background. It redirects the analytical focus from the citing author's intent to the actual knowledge contribution of the cited paper.
2) We propose a knowledge-centric annotation approach guided by a decision tree, and develop a Dual-Path Fusion model that integrates sentence-level semantics with document-level metadata. This model outperforms mainstream large language models (LLMs) and enables the automatic annotation of the entire ACL Anthology.
3) We apply the KCT to impact assessment and dissemination prediction, demonstrating that knowledge contribution classification offers a better basis for identifying high-impact papers and provides more stable predictive signals for scholarly dissemination than citation intent classification.

# 2. Related work

## 2.1 Citation Intent Classification

Citation intent classification aims to identify the functional role a citation serves for the citing paper, and represents one of the most mature research lines in citation content analysis. Research on citation intent is rooted in theoretical discussions of why researchers cite. Early studies showed that citation behavior is driven by multiple factors, including acknowledging intellectual debts, providing evidential support for arguments, identifying methodological sources, comparing results, and responding to peer criticism (Bornmann & Daniel, 2008; Tahamtan & Bornmann, 2019). Building on these motivations, the field has developed various intent taxonomies (Cohan et al., 2019; Duan & Tan, 2026; Lauscher et al., 2022; Teufel et al., 2006) among which the ACL-ARC framework by Jurgens et al. (2018) is the most representative, categorizing intents into Background, Use, Comparison, Extension, Motivation, and Future Work. Citation intent classification helps reveal rhetorical strategies in scholarly argumentation, deepens the understanding of interaction patterns in scientific communication, and provides a foundation for semantic enrichment of citation networks. These frameworks take the citing author's rhetorical motivation as the basis for classification, essentially characterizing the rhetorical function that a citation performs within a specific context.

Recent studies have attempted to improve the methods of intent classification using recent deep learning techniques. Ghosal et al. (2024) improved classification by incorporating cited paper title information. Quan et al. (2025) constructed a heterogeneous graph fusion network for multi-intent prediction. Koloveas et al. (2026) and Fogelson et al. (2025) explored the potential and reproducibility challenges of LLMs for intent classification. However, intent classification invariably targets the citing author's behavior and cannot provide a stable knowledge characterization of the cited paper. In contrast, the knowledge contribution classification in this study shifts the analytical focus to the cited paper itself, identifying the type of knowledge it contributes to the citing paper, thereby offering an alternative path to understanding citation behavior from the perspective of knowledge content.

## 2.2 Knowledge Contribution Classification

Identifying the knowledge contribution types of cited references requires establishing the theoretical premise that citations are not homogeneous. Important citation identification research addresses differences in the degree of citation influence, aiming to distinguish influential citations from peripheral ones. Teplitskiy et al. (2022) found that 54% of citations have “almost no influence” and primarily serve rhetorical purposes, providing direct empirical evidence for citation importance heterogeneity. Anderson and Lemken (2023) showed that citation context analysis can reveal which specific knowledge claims from a cited work are drawn upon by citing authors, often finding that a large proportion of citations are peripheral rather than substantive. These studies reveal the gap between citation counts and actual knowledge contributions but remain at the level of importance judgment, without further identifying what types of knowledge the cited papers contribute.

This gap calls for moving beyond measuring the degree of influence toward identifying the types of knowledge contributions. Existing studies have approached this issue from two perspectives. From a self-description perspective, Teufel and Moens (2002) identified author contribution statements through rhetorical status; Liakata et al. (2012) automatically recognized conceptualization zones in scientific articles; D'Souza et al. (2021) further extracted structured information units from papers to construct knowledge graphs; Pramanick et al. (2025) proposed an eight-category taxonomy and extracted author-stated contributions from a large-scale corpus of NLP paper abstracts to track their evolving trends. From a peer-absorption perspective, Chen et al. (2025) classified citation contexts to profile the contributions of Nature and Science papers and examined their relationship with author division of labor. Despite this progression, existing work has not yet established a formalized classification methodology to construct knowledge contribution taxonomies.

# 3. Methodology

## 3.1 Data Collection

We selected the ACL Anthology[1] as data source because it is the most comprehensive open-access repository of computational linguistics, offering full-text PDF access and standardized bibliographic metadata. A total of 84,722 full-text PDF documents spanning 45 years (1980–2024) were collected, and each was parsed into structured XML by deploying a local GROBID[2] service for batch processing. To extract in-text citation contexts, we developed a custom algorithm that captures a three-sentence window consisting of the citing sentence and its immediately preceding and following sentences, consistent with the approach of Zhang et al. (2022). This process yielded 1,131,743 raw in-text citation instances. We then applied a further data cleaning algorithm to remove non-English content, entries with missing or malformed citation markers, citation contexts dominated by code snippets, mathematical expressions, or tabular data, and citations referencing publications outside the ACL Anthology. After cleaning, 802,202 valid internal content citation records remained—where both the citing and cited papers belong to the ACL Anthology—covering 62,933 citing papers and 45,301 cited papers.

Figure 2 presents the statistical characteristics of the dataset. The yearly distribution exhibits exponential growth, reflecting the rapid development of the field. Figure 2(a) shows the temporal distribution across five periods corresponding to major paradigm shifts. As illustrated in Figure 2(b), citations appear primarily in the Introduction (44.0%) and Related Work (34.3%) sections. The citation context, defined as the citing sentence together with its immediately preceding and following sentences, exhibits a slightly right-skewed length distribution, with a median of 414 characters and a mean of 434 characters (Figure 2(c)).

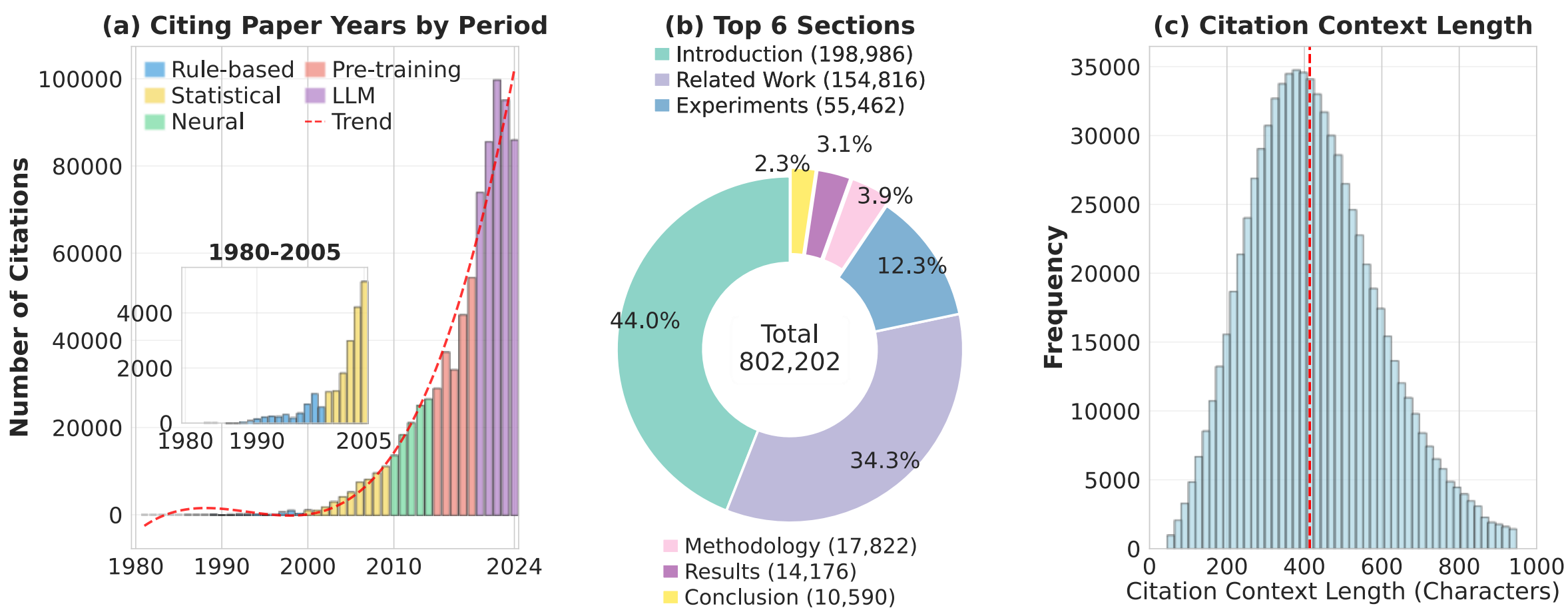

**Figure 2**. ACL Anthology Dataset Details

## 3.2 Knowledge Contribution Taxonomy

The KCT is grounded in the theoretical insight that citations are not homogeneous signals but carriers of specific types of knowledge. Small (1978) proposed that cited documents function as “concept symbols,” with each citation encoding a specific idea, method, or set of experimental data. Moravcsik and Murugesan (1975) provided early empirical evidence that citations differ not only in their importance but also in their intellectual function, distinguishing conceptual from operational and organic from perfunctory citations. However, not all citations carry equal knowledge weight. Tahamtan and Bornmann (2019) estimated that only 10–15% of citations perform substantive knowledge transfer functions. These observations call for a classification framework that distinguishes the types of knowledge that cited papers contribute.

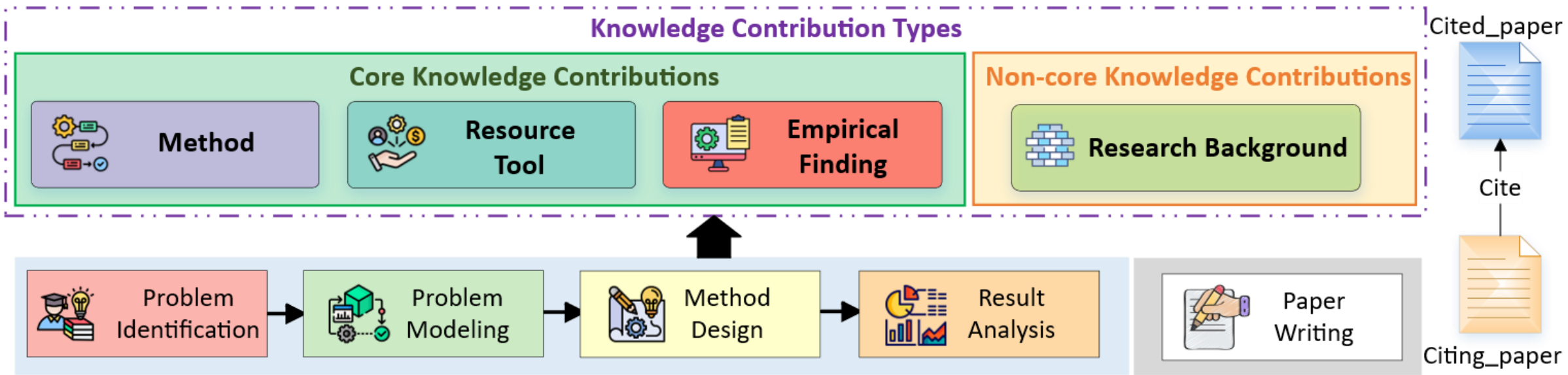

**Figure 3.** Scientific Research Logic Model

Drawing on the logical structure of the research process, we construct the Scientific Research Logic Model (Figure 3) that maps research activities onto a chain spanning problem identification, problem modeling, method design, and result analysis. The knowledge contribution classification framework is built upon this chain. The taxonomy categorizes the knowledge that cited papers contribute into two broad classes, namely Core Knowledge Contributions and Non-core Knowledge Contributions, encompassing four specific types (Table 1). Although the KCT was constructed and validated using data drawn entirely from the ACL Anthology, its classificatory logic is grounded in the general structure of the research process and thus holds potential for transfer to other domains.

**Table 1**. KCT Category Definitions

| Type | Definition | Example |
|---|---|---|
| **Method** | Citation where the citing paper directly uses, implements, improves, or extends methods, techniques, algorithms, models, systems, or evaluation metrics from the cited paper. Must be used by "we" (the authors). | We compute COMET scores [CITATION] separately for each domain with default wmt20-comet-da similarly to Table 4 |
| **Resource Tool** | Citation where the citing paper uses datasets, corpora, or explicitly labeled toolkits/toolboxes created by the cited paper. Must be directly used by "we" (the authors). | We first extracted opinionated and objective texts from DeReKo corpus [CITATION] |
| **Empirical Finding** | Citation providing specific, verifiable empirical finding (experimental results, performance data, observed phenomena) used for: 1) direct comparison, 2) justifying decisions, or 3) stating empirical facts. | Table 2 shows that, in the perspective of end-to-end discourse parsing, our parser first outperforms the state-of-the-art segmentator of [CITATION] |
| **Background** | Citation providing a necessary understanding foundation or positioning research within the field through: 1) developmental course (vertical timeline) or 2) research landscape (horizontal snapshot). Descriptive rather than operational. | Contemporary MT evaluation measures have evolved beyond simple lexical matching, and now take into account various aspects of linguistic structures [CITATION1]，[CITATION2]，[CITATION3] |

## 3.3 Expert Annotation

Full-scale annotation of the 802,202 citation contexts is infeasible. Since the citation data exhibit exponential growth (Figure 2(a)), simple random sampling would over-represent recent years. We therefore applied proportional stratified sampling by temporal period, drawing 2,000 instances whose distribution matches that of the full corpus, a scale comparable to related work in citation classification (Jurgens et al., 2018; Teufel et al., 2006). The sampled instances were annotated according to the KCT, with the annotation scheme and execution process detailed in the following subsections.

### 3.3.1 Annotation Scheme

This annotation scheme adopts a knowledge content priority principle, under which annotators identify the type of knowledge that the cited paper contributes rather than the citing author's intent. Unlike citation intent, which varies with the citing context, knowledge contribution classification is anchored to what the cited paper contributes, offering greater objectivity. To ensure accurate judgment, annotators are required to consider both the citation context and the cited paper's title and abstract. The annotation process follows the hierarchical decision tree in Figure 4, with criteria for each decision node defined in Table 2.

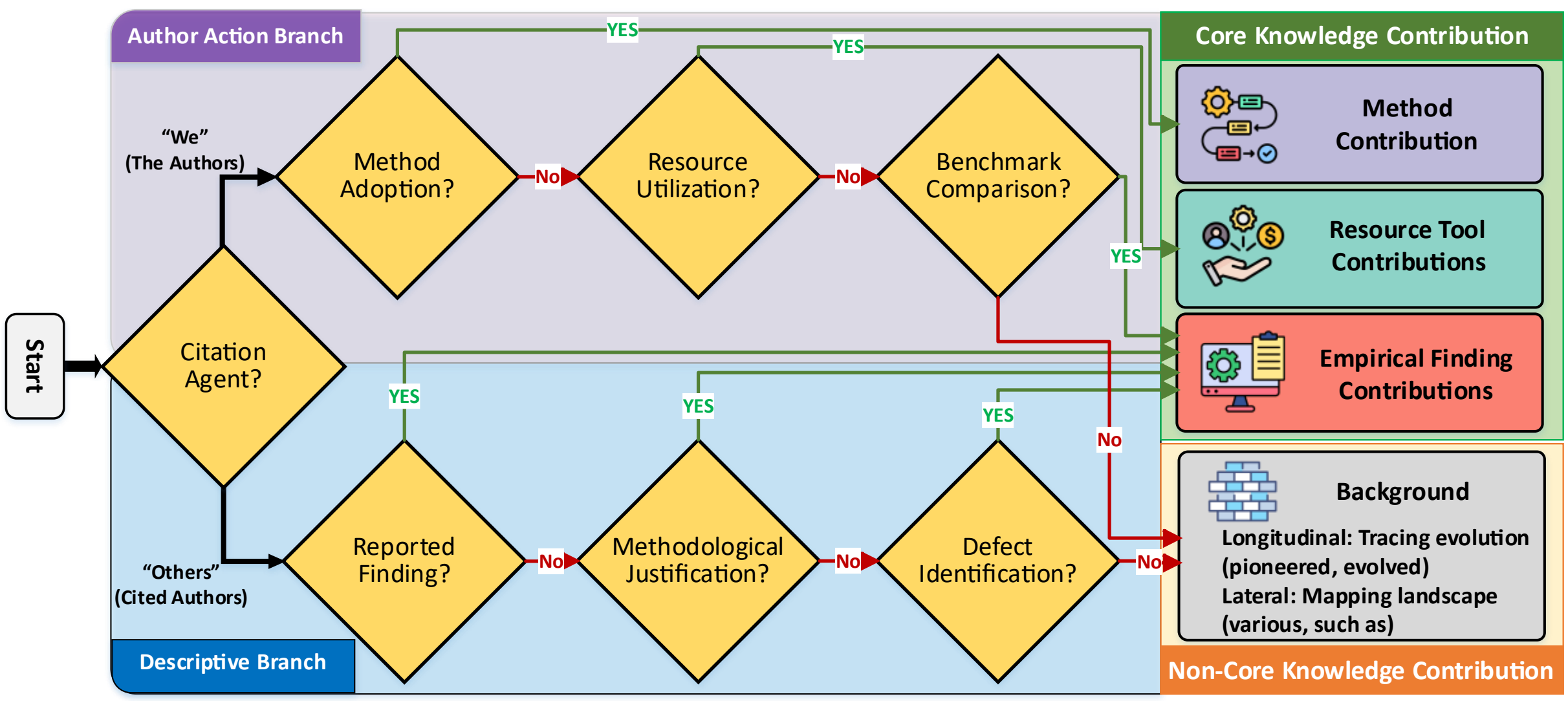

**Figure 4.** Hierarchical Classification Process for Knowledge Contribution

**Table 2.** Decision Nodes and Annotation Criteria of the Classification Decision Tree

| Branch | Decision Node | Detailed Criterion Explanation |
|---|---|---|
| Core Routing | Citation Agent? | Determines whether the agent of the action is the current authors (We) or the cited authors / objective entities (Others). For compound sentences, the agent of the clause directly containing the citation is used (proximity principle). |
| Author Action | Method Adoption? | Whether methods, techniques, algorithms, models, evaluation metrics, or systems from the cited work are directly adopted as the operational basis of the current study. |
| | Resource Utilization? | Whether datasets, corpora, or tools explicitly labeled as a toolkit/toolbox from the cited work are directly used. |
| | Benchmark Comparison? | Whether the cited work's performance or characteristics are cited as a comparison baseline to position the current work or highlight differences. |
| Descriptive | Reported Finding? | Whether the cited work uses a reporting verb (e.g., show, find, prove, report) to present concrete, verifiable empirical content, such as a numeric result or a definite conclusion, as opposed to merely describing what a work produced (e.g., propose, develop) or offering a macro-level value judgment. |
| | Methodological Justification? | Whether a finding or premise from the cited work is used as an explicit rationale for a specific research decision, forming a cited-finding-to-author-action chain. |
| | Defect Identification? | Whether a concrete defect, failure, or conflict in the cited work is identified and immediately used as the rationale for the authors' subsequent action, forming a problem-to-solution chain. |
| Final Class | Background | Citations satisfying no node within their branch. Their function is contextualization and positioning, typically tracing development (e.g., first, pioneered, evolved) or mapping the landscape (e.g., various, such as, multiple co-citations). |

The classification follows a two-branch structure determined by the subject of the citation sentence. When the current authors serve as the subject, the citation is routed to the author-action branch and evaluated sequentially against the criteria for Method, Resource Tool, and Empirical Finding. When other entities serve as the subject, the citation is routed to the descriptive branch and evaluated against three nodes, namely Reported Finding, Methodological Justification, and Defect Identification. A citation satisfying any one of these nodes receives an Empirical Finding label. This Dual-Pathway design ensures that empirical knowledge is captured whether it appears as active benchmarking or as a factual premise. Citations satisfying no criterion within their respective branch are classified as Background. The complete decision procedure, detailed definitions of all categories, and numbered annotation examples are provided in Section A of the supplementary material.

### 3.3.2 Data Annotation

Annotation was completed by four doctoral students in information science, each with over three years of experience in literature analysis. The annotators received rigorous training in the taxonomy and the annotation guidelines, and were calibrated on 100 pre-annotated samples to align their understanding. The annotation followed a two-stage process. In the first stage, 300 instances were jointly annotated by all four annotators to assess agreement, while the remaining 1,700 instances underwent cross-annotation, with each instance independently completed by at least two annotators. In the second stage, disagreements were resolved through group discussion among the annotators, with unresolved cases adjudicated by an expert outside the annotation team. An early version of the KCT included a fifth type for theoretical contributions. The annotation of the 2,000 instances, however, yielded only 3 cases (0.15%) under this type. This outcome reflects the characteristics of ACL as a technology-driven field, in which research contributions are manifested primarily in methodological innovation, resource construction, and empirical findings, rendering standalone theoretical contributions rare. This empirical finding motivated the consolidation of the taxonomy into its current four-type structure.

**Table 3.** Knowledge Contribution Type Distribution in Annotated Dataset

| Type | Count | Percentage |
|---|---|---|
| Method | 308 | 15.4% |
| Empirical Finding | 260 | 13.0% |
| Resource Tool | 124 | 6.2% |
| Background | 1308 | 65.4% |

Inter-annotator agreement was assessed over the 300 jointly annotated instances. The overall Fleiss' κ among the four annotators was 0.775, with a 95% confidence interval of [0.727, 0.822] and significantly above zero ($p < 0.001$), indicating that this level of agreement is statistically robust. By the benchmark of Landis and Koch (1977), this corresponds to substantial agreement. This level is comparable to or slightly higher than that of recent citation-annotation studies (Lauscher et al., 2022; Pramanick et al., 2025). At the class level, the one-vs-rest Cohen's κ averaged across annotator pairs is 0.88 for Method, 0.71 for Resource Tool, 0.67 for Empirical Finding, and 0.79 for Background, with all four reaching at least moderate agreement and the lowest value for Empirical Finding, consistent with its finer semantic boundary.

### 3.4 Automatic Annotation

#### 3.4.1 Model Architecture

The input feature set $\mathcal{F} = \{f_1, f_2, f_3, f_4\}$consists of four features across two complementary dimensions. The semantic dimension comprises the preceding sentence ($f_1$) and the current sentence ($f_2$), which capture fine-grained contextual information around the citation. The metadata dimension comprises the citing paper title ($f_3$) and the cited paper abstract ($f_4$), which provide macro-level literature information.

For the encoding strategy, we designed a Dual-Path architecture based on SciBERT to fully exploit complementary information between features. The first path adopts a Joint Encoding strategy, concatenating all features through special separator tokens [SEP] into a unified sequence and utilizing Transformer's self-attention mechanism to model cross-feature dependencies:

$$h_{joint} = \text{SciBERT}(f_1 \oplus [SEP] \oplus f_2 \oplus [SEP] \oplus f_3 \oplus [SEP] \oplus f_4) \quad (1)$$

where $h_{joint} \in \mathbb{R}^{768}$ is the pooled output of the joint sequence, $\oplus$ denotes sequence concatenation, and 768 is the hidden dimension of SciBERT.

The second path employs an Independent Encoding strategy, where each feature is separately encoded through SciBERT followed by mean pooling to preserve the independent semantic integrity of each feature:

$$h_i = \text{SciBERT}(f_i), i = 1, 2, \ldots, n \quad (2)$$

$$h_{indep} = \frac{1}{n}\sum_{i=1}^{n} h_i \quad (3)$$

where $h_i \in \mathbb{R}^{768}$is the pooled representation of the $i$-th feature, $h_{indep} \in \mathbb{R}^{768}$is the mean of the $n$ feature vectors, and $n = 4$ in this study. Independent encoding avoids information interference during feature mixing, preserving the fine-grained semantics of each feature.

We adopt a Hybrid Fusion strategy as the final architecture, integrating complementary information by concatenating the outputs of both paths:

$$z = h_{joint} \oplus h_{indep} \in R^{1536} \quad (4)$$

$$y = \text{softmax}(Wz + b) \quad (5)$$

where $W \in \mathbb{R}^{4\times1536}$, $b \in \mathbb{R}^4$ are learnable parameters corresponding to the four knowledge contribution categories. The advantage of hybrid fusion lies in: $h_{joint}$ captures inter-feature interaction patterns, $h_{indep}$ preserves the independent discriminative power of each feature, and the concatenation of dual paths avoids information loss from a single fusion strategy, enabling the model to understand citation knowledge contribution from both global and local perspectives simultaneously. We refer to this complete architecture as the Dual-Path Fusion model.

To address the class imbalance problem in knowledge contribution classification, we adopt Focal Loss as the training objective. Let $\gamma$ be the focusing parameter, the loss function is defined as:

$$\mathcal{L}_{focal} = -\frac{1}{N}\sum_{i=1}^{N} \alpha_{y_i}\left(1 - p_{y_i}\right)^{\gamma \log} p_{y_i} \tag{6}$$

where $N$ is the number of samples, $y_i$ is the true label for sample $i$, $p_{y_i}$ is the model's predicted probability for the true class, and $\alpha_{y_i}$ is the class weight (calculated as the inverse proportion of class frequency in the training set).

### 3.4.2 Experimental Setup

The final feature combination was determined through a two-stage preliminary experiment as shown in Section B of supplementary material. In the first stage, 1,586 valid combinations of 12 candidate features were ranked by macro-F1 under uniform default conditions for initial screening. In the second stage, the top 10 combinations were further optimized via grid search over fusion strategies and loss functions to fully exploit the classification potential of each combination. After this two-stage evaluation, the four-feature combination (preceding sentence + citing sentence + citing paper title + cited paper abstract) achieved the highest score under hybrid fusion with Focal Loss, indicating a stronger synergy between this combination and the advanced fusion and loss settings.

We used a fixed split of 1,500 training, 200 validation, and 300 test instances, with all metrics reported on the test set. Evaluation metrics include Accuracy and macro-averaged Precision, Recall, and F1-score. Given class imbalance, macro-averaging assigns equal weight to each category, providing a fairer assessment of model performance on minority classes. All experiments were conducted on a Linux server equipped with an NVIDIA GeForce RTX 4090 GPU with 24GB VRAM. The Dual-Path Fusion model was trained for 15 epochs with a batch size of 16, using the AdamW optimizer at a learning rate of 2e-5, linear warmup over 10% of total steps, and early stopping with a patience of 3. The $\gamma$parameter of Focal Loss was tuned on the validation set and set to 1.3. A few mainstream large language models were evaluated as comparison baselines, with inference performed via API calls under a zero-shot and a few-shot setting. The zero-shot setting used the task instruction alone. The few-shot setting added four representative examples to the prompt. The examples were drawn from the training set and covered all four categories. The temperature was set to 0.7 in both settings. All models were evaluated on the same 300-instance test set.

# 4. Results and Analysis

## 4.1 Classification Performance

To evaluate the performance of automatic knowledge contribution classification, we constructed a comparative framework spanning three paradigms as shown in Table 4. For traditional machine learning, we selected XGBoost (Chen & Guestrin, 2016). For Transformer-based pre-trained language models, we evaluated BERT (Devlin et al., 2019), SPECTER (Cohan et al., 2020), SciNCL (Ostendorff et al., 2022), and SciBERT (Beltagy et al., 2019). For LLMs, we evaluated DeepSeek-R1 (Guo et al., 2025), DeepSeek-V3.1 (DeepSeek-AI, 2025),

GPT-5-mini (OpenAI, 2025), Claude Sonnet 4.5 (Anthropic, 2025), Gemini 2.5 Flash (Comanici et al., 2025), and QwQ-32B (Qwen Team, 2025) under both zero-shot and few-shot prompting.

**Table 4.** Performance Comparison on Classification Task

| Model | ACC | Macro P | Macro R | Macro F1 |
|---|---|---|---|---|
| XGBoost | 0.685 | 0.349 | 0.353 | 0.353 |
| BERT | 0.805 | 0.723 | 0.703 | 0.703 |
| SPECTER | 0.816 | 0.709 | 0.721 | 0.714 |
| SciNCL | 0.835 | 0.735 | 0.738 | 0.738 |
| SciBERT | 0.847 | 0.757 | 0.757 | 0.755 |
| Gemini 2.5 Flash | 0.730/0.763 | 0.729/0.750 | 0.735/0.762 | 0.695/0.720 |
| DeepSeek-R1 | 0.797/0.790 | 0.720/0.727 | 0.738/0.731 | 0.724/0.721 |
| GPT-5-mini | 0.810/0.807 | 0.724/0.724 | 0.766/0.754 | 0.738/0.727 |
| DeepSeek-V3.1 | 0.780/0.813 | 0.710/0.753 | 0.675/0.714 | 0.678/0.728 |
| Claude Sonnet 4.5 | 0.743/0.820 | 0.644/0.726 | 0.789/0.749 | 0.685/0.734 |
| QwQ-32B | 0.770/0.797 | 0.738/0.740 | 0.700/0.761 | 0.705/0.740 |
| Dual-Path Fusion (ours) | **0.855** | **0.781** | **0.770** | **0.774** |

Note. LLM results are shown as zero-shot/few-shot and ordered by macro F1 of the few-shot strategy.

As shown in Table 4, XGBoost achieves an accuracy of 0.685 but a macro F1 of only 0.353. This gap reflects accuracy inflation under class imbalance and shows that the task needs deep semantic understanding beyond handcrafted features. Among the pre-trained language models, performance rises with the alignment between the pre-training signal and sentence-level classification. General-purpose BERT, with little exposure to scientific language, yields the lowest macro F1. SPECTER and SciNCL are already adapted to the scientific domain, yet their citation-based contrastive objectives target document-level similarity, so they improve only modestly. The SciBERT baseline attains the highest macro F1 with balanced Precision and Recall, because it is pre-trained from scratch on a scientific corpus with an in-domain vocabulary, and its token-level masked-language-modeling objective fits sentence-level discrimination better than document-embedding objectives. Therefore, we adopt it as the encoding backbone.

QwQ-32B, Claude Sonnet 4.5, and GPT-5-mini surpass BERT and SPECTER in terms of macro F1. Their broad pre-training already captures much of the semantics this task requires. Yet Recall stays well above Precision across the models, most visibly for Claude Sonnet 4.5 under zero-shot at 0.789 versus 0.644. Even at its best across the two settings, the strongest model reaches only 0.740 in macro F1 and still trails the SciBERT baseline. Domain-adapted pre-training therefore outweighs raw scale on this fine-grained task. Our Dual-Path Fusion model uses SciBERT as the encoding backbone and models the citing context and document-level metadata in parallel. The joint path models cross-feature dependencies, while the independent path preserves the fine-grained semantics of each feature. Fusing the two paths lifts macro F1 to 0.774, the best result among all compared models. An ablation on the four input features shows that they contribute unequally. The citing sentence ($f_2$) is the decisive component, since removing it collapses the model. Each of the remaining three features contributes only marginally. Performance varies across the four categories, yet the model stays effective on every class. Full ablation and per-class results are reported in Section C of the supplementary material.

## 4.2 Distribution of Knowledge Contributions

The distribution of knowledge contribution types across the annotated citations is markedly uneven. Background dominates the corpus with 488,618 instances (60.91%), while Method (141,671, 17.66%), Empirical Finding (118,325, 14.75%), and Resource Tool (53,587, 6.68%) together account for less than 40%. This means that over 60% of citation links in conventional citation analysis do not carry direct knowledge contributions, underscoring the necessity of distinguishing knowledge contribution types to accurately characterize knowledge flow.

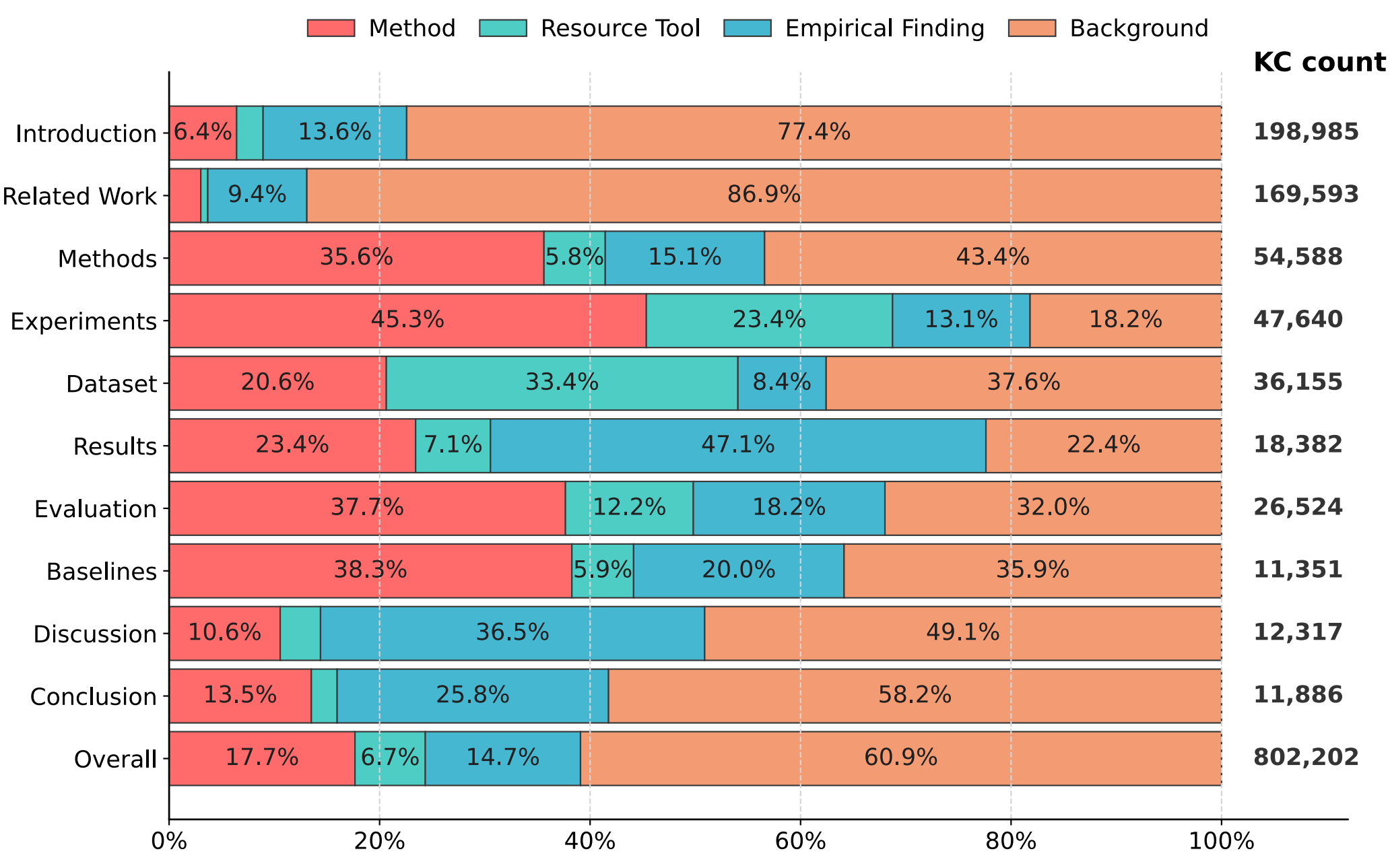

**Figure 5** Distribution of Knowledge Contribution Types Across Paper Narrative Structure

The distribution of knowledge contribution types across the paper narrative structure exhibits notable section-level variation (Figure 5). In Introduction and Related Work, which serve contextual framing functions, Background dominates overwhelmingly, indicating that citations in these sections are primarily used to establish the research context and survey prior work. Moving into Methods and Experiments, the distribution shifts substantially, with Method rising to prominence, reflecting the concentrated demand for referencing specific technical approaches and experimental tools in these sections. In Results, Empirical Finding reaches its highest proportion across all sections, corresponding to comparisons and corroborations between the reported results and existing empirical evidence. This alignment between knowledge contribution types and the argumentative functions of different sections indicates that citation behavior is a knowledge selection process guided by the argumentative objectives of each section, which also validates the KCT from an applied perspective.

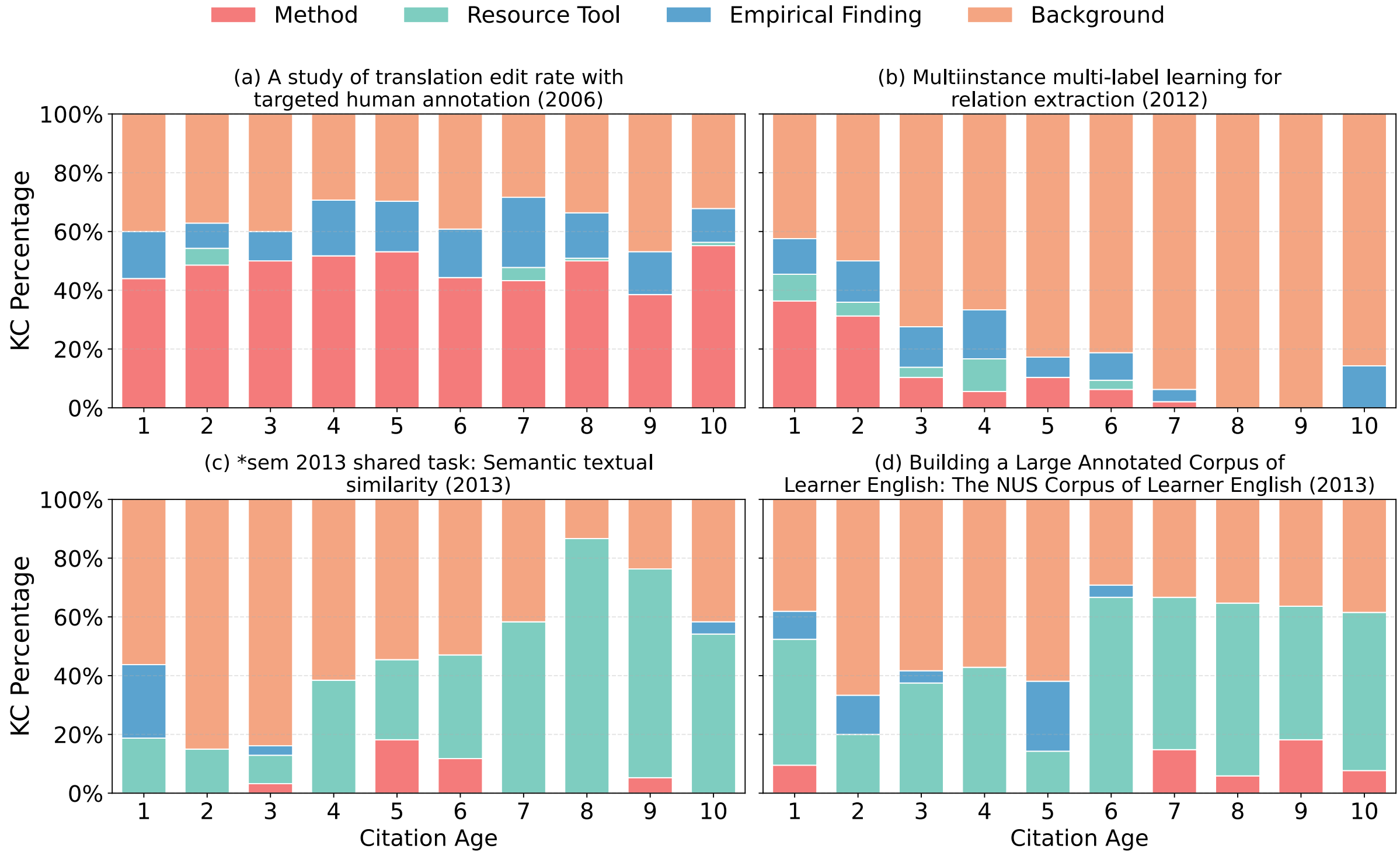

**Figure 6**. Temporal Evolution of Knowledge Contribution Types in Highly Cited Papers

To illustrate how KCT operates at the individual paper level, we selected four representative cases. From papers published between 2005 and 2014 with sufficient annual citations across citation ages 1 to 10, we computed JSD between the early-phase (age 1–5) and late-phase (age 6–10) knowledge contribution distributions and identified four evolutionary trajectories (Figure 6). Paper (a), with JSD = 0.035, maintains a stable distribution, with Method consistently around 46–50%. Paper (b), with JSD = 0.334, undergoes a pronounced decline in Method accompanied by a rise in Background to 91.1%. Paper (c), with JSD = 0.396, shows the opposite trend, with Resource Tool surging to 62.3%. Paper (d), with JSD = 0.308, shifts in dominant type from Background to Resource Tool. These trajectories demonstrate that a single paper may carry different knowledge contributions across its lifespan, a dynamic differentiation invisible to citation counts alone.

## 4.3 Impact Evaluation Based on Knowledge Contributions

In traditional impact assessment, every citation is treated as an equal unit, ignoring knowledge contribution differences among citations. This section examines whether excluding non-core knowledge contribution citations can improve the identification of high-impact papers. To this end, we designed an award-winning paper recognition task. We extracted award-winning paper records from award tags in the ACL Anthology XML metadata and from the ACL Wiki Best Paper Awards page[3] respectively. After merging and deduplication, 174 award-winning papers were matched, covering four categories: Best Paper, Outstanding Paper, Test of Time, and Best Resource. Then, we constructed two rankings for all 45,301 cited papers, one based on raw citation count and the other based solely on core knowledge contribution citation count, and compared hit rates for award-winning papers at different cutoffs. At all cutoffs, the core knowledge contribution citation count achieves higher hit rates than raw citation count, with improvements of 2.7% to 6.7% (Table 5). At Top 75%, it captures all 174 award-winning papers while raw citation count still misses 11, indicating that non-core knowledge contribution citations dilute the discriminative power of traditional citation counts.

We further attempted to confirm the advantage of core knowledge contribution citations in distinguishing award-winning papers. We matched each of the 174 award-winning papers to non-award papers from the same venue, with publication year difference less than 2 and a comparable citation count. A conditional logistic regression analysis showed that the share of core knowledge contributions could be a better predictor for award-winning papers (odds ratio 3.53, 95% CI 1.80 to 6.92, $p<0.001$) than citation count. Detailed results can be found in Section D of the supplementary material.

**Table 5.** Hit Rates of Award-Winning Papers under Raw and Core Citation Counts

| Ranking cutoff | Citation Count | | Core knowledge contribution citation count | | | |
|---|---|---|---|---|---|---|
| | **Hits** | **Rate** | **Hits** | **Rate** | **Δ Hits** | **Δ (%)** |
| 5% | 55 | 31.61% | 58 | 33.33% | +3 | +5.5% |
| 10% | 74 | 42.53% | 76 | 43.68% | +2 | +2.7% |
| 25% | 111 | 63.79% | 117 | 67.24% | +6 | +5.4% |
| 50% | 140 | 80.46% | 146 | 83.91% | +6 | +4.3% |
| 75% | 163 | 93.68% | 174 | 100.00% | +11 | +6.7% |

**Note:** Hits is the number of award papers captured within the top N% ranked by each indicator. Rate is Hits divided by 174 times 100%. Δ(%) is Core KC Hits minus Raw Hits, divided by Raw Hits, times 100%. Ties are resolved inclusively, so all papers with citation counts at or above the cutoff score at each threshold are included.

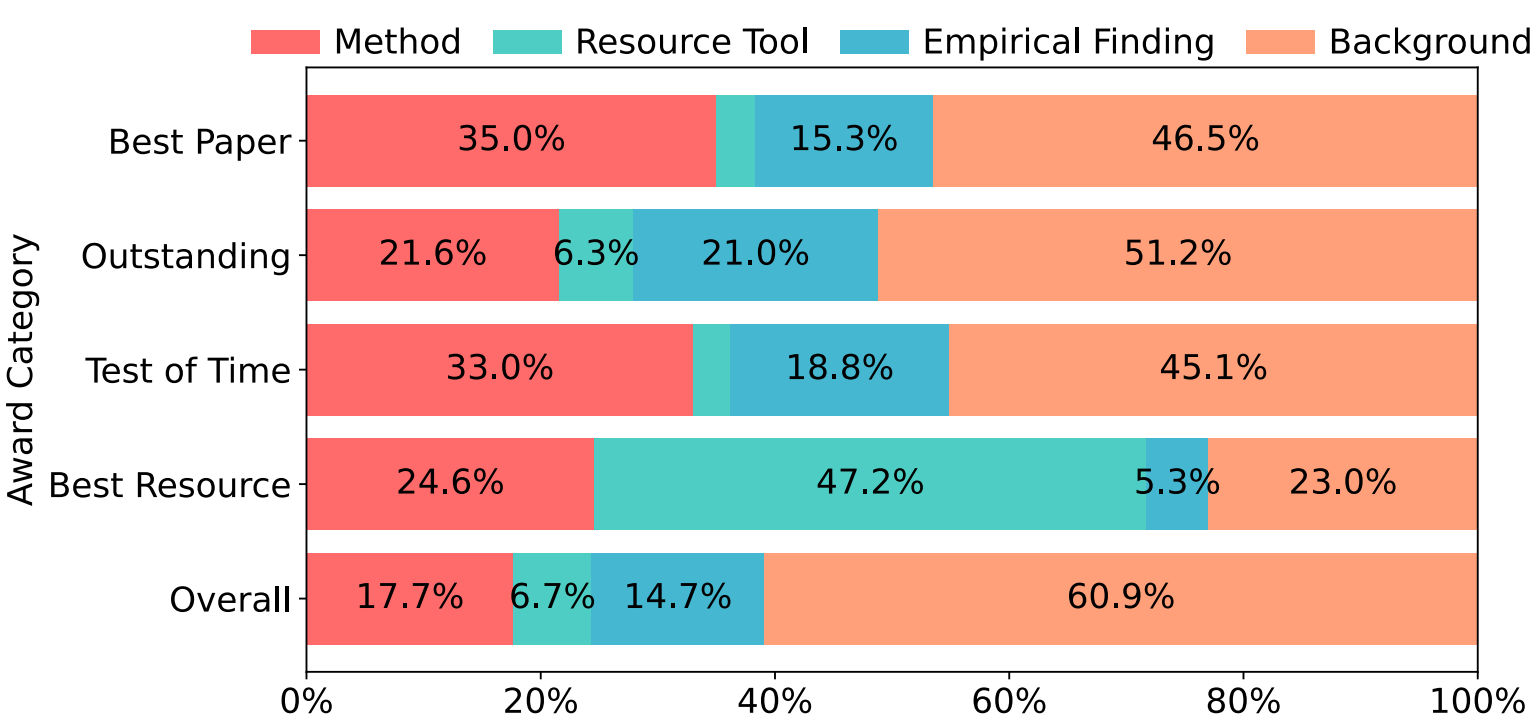

**Figure 7.** Knowledge Contribution Composition of Award-Winning Papers across Award Categories

We decomposed the citations of award-winning papers by knowledge contribution type (Figure 7). Across all four award categories, the Background share fell well below the overall mean of 60.9%. Award-winning papers were thus cited more for core contributions than papers in general. This is why a core-oriented indicator separated them. The leading core type, however, differed by category. Best Paper and Test of Time were strongest in Method, whereas Best Resource was strongest in Resource Tool, the latter reaching 47.2%, about seven times the overall mean. A raw citation count collapses these differences into a single number. The knowledge contribution types instead recover them as distinct dimensions of influence, which is what lets a core-oriented indicator identify high-impact papers beyond citation volume.

Different types of knowledge contributions are not equally relevant in impact evaluation, and their importance depends on the specific analytical purpose. For instance, identifying key resource tool hubs within a field requires Resource Tool citations as the core signal rather than aggregate citations (Bezerra et al., 2026). Following this rationale, we constructed directed citation sub-networks by retaining only Method, Resource Tool, and Empirical Finding citations respectively, preserving 21.6%, 9.0%, and 18.1% of edges in the full network.

Table 6 reports the in-degree rankings of the Top 10 papers within the three sub-networks, together with their core-contribution shares. Each sub-network retains citations of a single knowledge type. A paper can therefore rank very differently across dimensions, and these gaps show where its influence concentrates. ROUGE ranks 3rd overall but 75th as a Resource Tool, and ELMo ranks 7th but 181st. Both are cited as an evaluation metric or a baseline rather than as a reusable tool. BERT enters the Top 5 on all three dimensions, yet its 68.2% Method share shows that ranking breadth and influence balance are distinct properties. These sub-network ranks propose no new ordering. They decompose one citation count into separate kinds of influence. The divergence is not confined to a few prominent papers but holds network-wide. Across all papers, the Spearman correlation with the overall rank is only 0.393 for Resource Tool, against 0.628 for Method and 0.654 for Empirical Finding. The overall count therefore fails to capture resource-tool influence. Consistent with this, 55.4% of the Resource Tool Top 100 lie outside the overall Top 100. Impact evaluation should therefore read the dimension matched to its purpose rather than rely on aggregate counts alone.

**Table 6**. Sub-network Rankings and Core-Contribution Shares of the Top 10 Papers

| Paper | Overall Rank | Sub-network Rankings | | | Share of M/R/E (%) |
|---|---|---|---|---|---|
| | | Method | Resource Tool | Empirical Finding | |
| BERT | 1 | 1 | 4 | 1 | 68.2/7.7/24.0 |
| BLEU | 2 | 2 | 29 | 2 | 76.3/3.0/20.7 |
| ROUGE | 3 | 3 | 75 | 4 | 79.6/3.8/16.6 |
| Transformer | 4 | 9 | 1 | 191 | 30.8/67.5/1.7 |
| Moses | 5 | 4 | 2 | 32 | 55.6/38.6/5.7 |
| BART | 6 | 5 | 28 | 6 | 68.4/10.6/21.1 |
| ELMo | 7 | 17 | 181 | 3 | 55.8/4.8/39.4 |
| XLM-R | 8 | 10 | 19 | 5 | 57.4/16.3/26.3 |
| Sentence-BERT | 9 | 7 | 42 | 12 | 73.8/10.4/15.8 |
| GIZA++ | 10 | 6 | 23 | 36 | 77.1/14.1/8.9 |

**Note:** Ranks are based on in-degree (intra-ACL Anthology citation count), with ties assigned the minimum rank. M, R, and E denote Method, Resource Tool, and Empirical Finding, respectively.

## 4.4 KCT versus Citation Intent in Dissemination Prediction

Predicting the impact of research papers is an established line of work. This section compares two ways of characterizing citations, knowledge contribution classification and intent classification, for predicting scholarly dissemination, adopting the citation intent framework of Jurgens et al. (2018) as the comparative benchmark. Prior work often draws on a paper's own content. For example, Vital et al. (2025) use text embeddings to predict impact from a paper's text. We hold the opinion that the early citations of a paper may predict its dissemination. Thus, we investigate whether the knowledge contribution structure of early citations predicts later dissemination. We apply a heterogeneous graph fusion network-based multi-intent prediction algorithm (Quan et al., 2025) to annotate all citation records in the ACL Anthology with intent labels. On this basis, we use the label distribution of citations received within the first three years after each cited paper's publication as features to predict its scholarly dissemination performance over the subsequent seven years. The feature window and the prediction target are strictly separated in time. To ensure a fair comparison across papers published in different years, all observation windows are fixed at 10 years post-publication, and any citations received beyond this period are excluded. If the early-stage label distribution from one framework predicts later performance more accurately, this indicates stronger predictive validity for scholarly dissemination.

We construct four prediction tasks. 1) Citation Volume, measured by the natural logarithm of cumulative citation counts, captures the overall magnitude of impact accumulation. 2) Venue Diversity, measured by the Shannon entropy of citing venues, captures the breadth of cross-community dissemination. 3) Persistence, defined as whether a paper receives citations in at least five distinct years, captures whether knowledge maintains sustained vitality. 4) Backgrounding, defined as whether Background citations account for more than 50% of all citations, captures whether a cited paper gradually degenerates into background reference. The first two are continuous-value prediction tasks and the latter two are binary classification tasks. To rule out the influence of model selection on conclusions and ensure a robust and fair comparison between the two frameworks, each task employs both a linear and a nonlinear model. Continuous-value tasks use Ridge regression and a LightGBM regressor, while binary classification tasks use logistic regression and a LightGBM classifier. The sample retains only cited papers published no later than 2014 with at least five citations in the first three years, ensuring completeness of the observation period and estimation precision of early-stage label distributions. The final sample comprises 4,249 papers.

We set up three experimental conditions. Baseline includes six control variables, namely publication year, log of citation count in the first three years, log of author count, title word count, log of abstract word count, and a binary indicator for whether the paper was published at an ACL venue. Intent and KCT each augment Baseline by adding the proportional distribution of citation intent categories and knowledge contribution types, respectively. We employ nested 10-fold cross-validation in all experiments, where the outer 10 folds evaluate model performance and the inner 5 folds tune hyperparameters, with strict separation between the two layers to prevent information leakage. The primary metric is RMSE for continuous-value tasks and AUC for binary classification tasks. Statistical significance of differences between frameworks is determined via paired bootstrap tests with 1,000 resamples.

**Table 7.** Prediction Performance Using Linear Models

| Setting | Ridge Regression (RMSE↓) | | Logistic Regression (AUC ↑ ) | |
|---|---|---|---|---|
| | **Citation Volume** | **Venue Diversity** | **Persistence** | **Backgrounding** |
| Baseline | 1.034 | 0.617 | 0.740 | 0.575 |
| Intent | 1.025 | 0.612 | 0.752 | 0.689 |
| KCT | **1.018** | **0.607** | **0.755** | **0.707** |

**Table 8.** Prediction Performance Using Nonlinear Models

| Setting | LightGBM Regressor (RMSE↓) | | LightGBM Classifier (AUC↑) | |
|---|---|---|---|---|
| | **Citation Volume** | **Venue Diversity** | **Persistence** | **Backgrounding** |
| Baseline | 1.029 | 0.621 | 0.739 | 0.562 |
| Intent | 1.022 | 0.614 | 0.746 | 0.688 |
| KCT | **1.015** | **0.610** | **0.754** | **0.698** |

Note：↓ = lower is better, ↑ = higher is better.

Across both linear and nonlinear models, KCT predicts later dissemination more accurately than the citation intent framework (Tables 7 and 8). KCT outperforms Intent in all eight comparisons, and Intent never significantly exceeds KCT under the paired bootstrap test. For Citation Volume, the most basic impact measure, KCT significantly lowers the Ridge error from 1.025 to 1.018 ($p = 0.012$). The Backgrounding task is the clearest. Its Baseline AUC is only 0.575, close to random. Adding KCT raises it to 0.707 and significantly exceeds Intent ($p = 0.026$). A paper can thus keep accumulating citations after its substantive influence has faded, and KCT detects this shift earlier. The two frameworks differ in their informational orientation. Knowledge contribution types are constrained by the cited paper's content, so their early distribution already captures its contribution profile. A split-half consistency analysis confirms this stability. Resource Tool shows the highest cross-period correlation ($r = 0.545$), far above the other types (Figure E1 of the supplementary material). Citation intent, in contrast, shifts with research contexts and paradigms, so its early distribution carries weaker information about later dissemination. These results point to a broader implication for assessing dissemination. Citation counts keep rising even as a paper's substantive role decays, so volume alone overstates lasting influence. Tracking how the knowledge roles of citations shift over time gives an earlier and more faithful interpretation of scholarly dissemination than counting citations.

# 5. Discussion

## 5.1 From Citation Intent to Knowledge Contribution

Citation intent classification and knowledge contribution classification differ in analytical perspective. The former infers the citing author's subjective purpose, so the same paper may receive different intent labels across argumentative scenarios, making such judgments inherently uncertain. The latter is anchored in the knowledge content of the cited paper, offering a more objective basis. The dissemination prediction experiments support this. Across all comparisons over both linear and nonlinear models, KCT predicts more strongly than the citation intent framework, with multiple comparisons reaching significance. The early-stage distribution of knowledge contribution types thus already predicts later dissemination reliably, whereas intent labels, subject to shifting citing contexts, carry weaker early signal.

The deeper value of the knowledge contribution perspective is that it turns each citation into a traceable knowledge transmission behavior, so we know not only that a citation occurs, but also what type of knowledge is flowing. This study assigns each citation a single label, its dominant knowledge contribution type. A citation may in principle involve several types at once, for instance when a cited work supplies both a method and a dataset. We take the dominant type because our aim is to characterize the primary role a citation plays in knowledge transmission, and this role is what impact and dissemination analysis build on. Small (1978) conceptualized cited papers as concept symbols but did not distinguish the knowledge these symbols carry. KCT extends that concept into four semantically explicit dimensions, so citation networks become knowledge transmission networks in which the pathway of each knowledge type is independently analyzable.

The two frameworks closest to ours come from different fields, yet both ground their categories in corpus, whereas ours rest on a model of the research process. Pramanick et al. (2025), in NLP, read author statements in abstracts, so their categories describe what authors claim. Chen et al. (2025), in scientometrics, infer contributions at the paper level from aggregated citation contexts, whereas our unit is the individual citation. That two frameworks from different fields still differ from ours shows that categories follow from how they are constructed, not from the corpus.

Our categories are constructed from the Scientific Research Logic Model, a general account of empirical inquiry that decomposes any study into problem identification, problem modeling, method design and result analysis. These four stages are not specific to computational linguistics but describe the shared structure of scientific work. The ACL corpus serves only to instantiate the resulting types within ACL. This model-based origin is what makes the framework portable. When transferring it to a new discipline, a new framework of the categories can be derived from the same four stages, rather than reusing the ACL ones. A more theory-oriented discipline, for instance, may yield a Theoretical Contribution type. The process is therefore invariant while its output adapts to each field.

## 5.2 Toward Knowledge Contribution-Based Research Evaluation

Current research evaluation treats every citation as an equal edge. Yet papers with the same citation count may serve different knowledge contribution types. One response is to weight citations rather than treating them equally. For instance, studies have distinguished influential citations from incidental ones (Valenzuela et al., 2015) and have incorporated citation reputation into evaluation frameworks (Safón et al., 2025). Such weighting still measures the extent to which a citation matters, rather than the kind of knowledge it carries. Knowledge contribution classification shifts the evaluative focus from citation volume to the content of knowledge transfer. Across impact assessment and dissemination prediction, it surfaces signals that citation counts and citation intent miss. A citation's evaluative meaning thus depends on the knowledge it carries, not on its occurrence. The evidence is concrete. The award categories in our dataset show distinct contribution profiles (Figure 7). A highly cited paper often holds core status in only one knowledge dimension (Table 6). No single count can therefore serve every evaluative purpose.

The KCT enables peer review and funding agencies to recognize the contributions that are obscured by citation counts alone. Resource Tool citations are a case in point. They form only a small share of the citation network yet dominate a distinct influence dimension, so the builders of datasets and toolkits, whose own citation counts may be modest, are systematically undervalued whenever influence is read off a single number. The backgrounding results expose a second blind spot. A rising citation count is easily read as continuing influence, yet a paper's core contribution share can fall even as its citations keep accumulating. Evaluation should therefore ask not only how many citations a paper attracts but also how its contribution structure shifts over time.

This semantic lens can be extended in several directions. Integrating FAIR principles (Wilkinson et al., 2016) would let it weigh the accessibility and reuse of cited resources. It can also be analyzed in conjunction with intent classification. Knowledge contributions and citation intents are two facets that Small (1982) had already placed within a single program of citation context analysis, yet later context-based evaluation has tended to treat them separately. Combining them together would show both what knowledge a paper supplies and how that knowledge is used. Each direction points to the same underlying shift, a change in the unit of evaluation. The question moves from how often a work is cited to what it contributes and how that contribution matters. It is this shift, rather than any single new indicator, that allows research evaluation to engage with the substance of knowledge transfer.

### 5.3 Limitations

Several limitations of this study should be acknowledged. The analysis covers only the field of computational linguistics, and whether the KCT and he observed distributions of knowledge contribution types generalize to other disciplines requires further validation. Future work could transplant the process of constructing the KCT to other fields and develop discipline-specific classification schemes with automated annotation. The evaluation in our experiments kept the test set at its natural, imbalanced distribution and did not apply balanced test sets or resampling, which is a direction worth exploring in future work.

## 6. Conclusion

This study proposes the KCT, redirecting the analytical focus of citation analysis from the citing author's intent to the type of knowledge that the cited paper contributes. We develop a Dual-Path Fusion model that achieves 85.5% classification accuracy, outperforming mainstream LLMs. Large-scale analysis on the ACL dataset reveals that different knowledge types are distributed in clear correspondence with the argumentative structure of papers, and that the same paper may serve different knowledge contribution types at different stages of its lifecycle. For impact assessment, the core knowledge contribution citation count consistently outperforms traditional citation count in identifying award-winning papers. Citation sub-networks constructed by knowledge type further reveal ranking differences masked by aggregate counts. In dissemination prediction, KCT outperforms intent classification, demonstrating stronger predictive validity for scholarly dissemination. By shifting the unit of analysis from citation frequency to knowledge content, the KCT opens new possibilities for more fine-grained research evaluation, and provides methodological support for objectively characterizing knowledge transfer in citations.

# ACKNOWLEDGMENTS

This article was funded by the National Natural Science Foundation of China (Grant Nos. 72174154, 72504210).

# References

Aman, V., & Gläser, J. (2025). Investigating Knowledge Flows in Scientific Communities: The Potential of Bibliometric Methods. *Minerva*, *63*(1), 155–182. https://doi.org/10.1007/s11024-024-09542-2

Anderson, M. H., & Lemken, R. K. (2023). Citation Context Analysis as a Method for Conducting Rigorous and Impactful Literature Reviews. Organizational Research Methods, 26(1), 77–106. https://doi.org/10.1177/1094428120969905

Anthropic. (2025). Claude Sonnet 4.5 System Card [Technical Report]. Anthropic. https://www.anthropic.com/research/claude-sonnet-4-5-system-card

Beltagy, I., Lo, K., & Cohan, A. (2019). SciBERT: A Pretrained Language Model for Scientific Text (arXiv:1903.10676). arXiv. https://doi.org/10.48550/arXiv.1903.10676

Bezerra, D. A., Silva, F. N., & Amancio, D. R. (2026). Leveraging GANs for citation intent classification and its impact on citation network analysis. Journal of Informetrics, 20(2), 101791. https://doi.org/10.1016/j.joi.2026.101791

Bornmann, L., & Daniel, H. (2008). What do citation counts measure? A review of studies on citing behavior. Journal of Documentation, 64(1), 45–80. https://doi.org/10.1108/00220410810844150

Bornmann, L., & Leibel, C. (2026). Citation accuracy, citation noise, and citation bias: A foundation of citation analysis (arXiv:2508.12735). arXiv. https://doi.org/10.48550/arXiv.2508.12735

Chen, L., Ding, J., Song, D., & Qu, Z. (2025). Exploring scientific contributions through citation context and division of labor. Scientometrics, 130(5), 2901–2921. https://doi.org/10.1007/s11192-025-05318-x

Chen, T., & Guestrin, C. (2016). XGBoost: A Scalable Tree Boosting System. Proceedings of the 22nd ACM SIGKDD International Conference on Knowledge Discovery and Data Mining, 785–794. https://doi.org/10.1145/2939672.2939785

Cohan, A., Ammar, W., van Zuylen, M., & Cady, F. (2019). Structural Scaffolds for Citation Intent Classification in Scientific Publications. In J. Burstein, C. Doran, & T. Solorio (Eds.), Proceedings of the 2019 Conference of the North American Chapter of the Association for Computational Linguistics: Human Language Technologies, Volume 1 (Long and Short Papers) (pp. 3586–3596). Association for Computational Linguistics. https://doi.org/10.18653/v1/N19-1361

Cohan, A., Feldman, S., Beltagy, I., Downey, D., & Weld, D. S. (2020). SPECTER: Document-level Representation Learning using Citation-informed Transformers (arXiv:2004.07180). arXiv. https://doi.org/10.48550/arXiv.2004.07180

Comanici, G., Bieber, E., Schaekermann, M., Pasupat, I., Sachdeva, N., Dhillon, I., Blistein, M., Ram, O., Zhang, D., Rosen, E., Marris, L., Petulla, S., Gaffney, C., Aharoni, A., Lintz, N., Pais, T. C., Jacobsson, H., Szpektor, I., Jiang, N.-J., … Helmholz, W. (2025). Gemini 2.5: Pushing the Frontier with Advanced Reasoning, Multimodality, Long Context, and Next Generation Agentic Capabilities (arXiv:2507.06261). arXiv. https://doi.org/10.48550/arXiv.2507.06261

DeepSeek-AI. (2025). DeepSeek-V3.1. Hugging Face. https://huggingface.co/deepseek-ai/DeepSeek-V3.1

Devlin, J., Chang, M.-W., Lee, K., & Toutanova, K. (2019). BERT: Pre-training of Deep Bidirectional Transformers for Language Understanding. In J. Burstein, C. Doran, & T. Solorio (Eds.), Proceedings of the 2019 Conference of the North American Chapter of the Association for Computational Linguistics: Human Language Technologies, Volume 1 (Long and Short Papers) (pp. 4171–4186). Association for

Computational Linguistics. https://doi.org/10.18653/v1/N19-1423

D'Souza, J., Auer, S., & Pedersen, T. (2021). SemEval-2021 Task 11: NLPContributionGraph - Structuring Scholarly NLP Contributions for a Research Knowledge Graph. In A. Palmer, N. Schneider, N. Schluter, G. Emerson, A. Herbelot, & X. Zhu (Eds.), Proceedings of the 15th International Workshop on Semantic Evaluation (SemEval-2021) (pp. 364–376). Association for Computational Linguistics. https://doi.org/10.18653/v1/2021.semeval-1.44

Duan, C., & Tan, Z. (2026). Semantically Orthogonal Framework for Citation Classification: Disentangling Intent and Content. In W.-T. Balke, K. Golub, Y. Manolopoulos, K. Stefanidis, & Z. Zhang (Eds.), Linking Theory and Practice of Digital Libraries (pp. 183–206). Springer Nature Switzerland. https://doi.org/10.1007/978-3-032-05409-8_12

Erikson, M. G., & Erlandson, P. (2014). A taxonomy of motives to cite. Social Studies of Science, 44(4), 625–637. https://doi.org/10.1177/0306312714522871

Fogelson, A., Trišović, A., & Thompson, N. (2025). LLMs in Citation Intent Classification: Progress, Precision, and Reproducibility Challenges. Proceedings of the 3rd ACM Conference on Reproducibility and Replicability, ACM REP '25, 250–253. https://doi.org/10.1145/3736731.3746137

Garfield, E. (1955). Citation Indexes for Science: A New Dimension in Documentation through Association of Ideas. Science, 122(3159), 108–111. https://doi.org/10.1126/science.122.3159.108

Ghosal, T., Varanasi, K. K., & Kordoni, V. (2024). A Deep Multi-Tasking Approach Leveraging on Cited-Citing Paper Relationship For Citation Intent Classification. Scientometrics, 129(2), 767–783. https://doi.org/10.1007/s11192-023-04811-5

Guo, D., Yang, D., Zhang, H., Song, J., Wang, P., Zhu, Q., Xu, R., Zhang, R., Ma, S., Bi, X., Zhang, X., Yu, X., Wu, Y., Wu, Z. F., Gou, Z., Shao, Z., Li, Z., Gao, Z., Liu, A., … Zhang, Z. (2025). DeepSeek-R1 incentivizes reasoning in LLMs through reinforcement learning. Nature, 645(8081), 633–638. https://doi.org/10.1038/s41586-025-09422-z

Jurgens, D., Kumar, S., Hoover, R., McFarland, D., & Jurafsky, D. (2018). Measuring the Evolution of a Scientific Field through Citation Frames. Transactions of the Association for Computational Linguistics, 6, 391–406. https://doi.org/10.1162/tacl_a_00028

Koloveas, P., Chatzopoulos, S., Vergoulis, T., & Tryfonopoulos, C. (2026). Can LLMs Predict Citation Intent? An Experimental Analysis of In-Context Learning and Fine-Tuning on Open LLMs. In W.-T. Balke, K. Golub, Y. Manolopoulos, K. Stefanidis, & Z. Zhang (Eds.), Linking Theory and Practice of Digital Libraries (pp. 207–224). Springer Nature Switzerland. https://doi.org/10.1007/978-3-032-05409-8_13

Kunnath, S. N., Herrmannova, D., Pride, D., & Knoth, P. (2021). A meta-analysis of semantic classification of citations. Quantitative Science Studies, 2(4), 1170–1215. https://doi.org/10.1162/qss_a_00159

Kunnath, S. N., Pride, D., & Knoth, P. (2023). Prompting Strategies for Citation Classification. Proceedings of the 32nd ACM International Conference on Information and Knowledge Management, CIKM '23, 1127–1137. https://doi.org/10.1145/3583780.3615018

Landis, J. R., & Koch, G. G. (1977). The Measurement of Observer Agreement for Categorical Data. Biometrics, 33(1), 159. https://doi.org/10.2307/2529310

Lauscher, A., Ko, B., Kuehl, B., Johnson, S., Cohan, A., Jurgens, D., & Lo, K. (2022). MultiCite: Modeling realistic citations requires moving beyond the single-sentence single-label setting. Proceedings of the 2022 Conference of the North American Chapter of the Association for Computational Linguistics: Human Language Technologies, 1875–1889. https://doi.org/10.18653/v1/2022.naacl-main.137

Liakata, M., Saha, S., Dobnik, S., Batchelor, C., & Rebholz-Schuhmann, D. (2012). Automatic recognition of conceptualization zones in scientific articles and two life science applications. Bioinformatics, 28(7), 991–1000. https://doi.org/10.1093/bioinformatics/bts071

Moravcsik, M. J., & Murugesan, P. (1975). Some Results on the Function and Quality of Citations. Social Studies of Science, 5(1), 86–92. https://doi.org/10.1177/030631277500500106

OpenAI. (2025). GPT-5 System Card [Technical Report]. OpenAI. https://openai.com/index/gpt-5-system-card/

Ostendorff, M., Rethmeier, N., Augenstein, I., Gipp, B., & Rehm, G. (2022). Neighborhood Contrastive Learning for Scientific Document Representations with Citation Embeddings (arXiv:2202.06671). arXiv. https://doi.org/10.48550/arXiv.2202.06671

Pramanick, A., Hou, Y., Mohammad, S. M., & Gurevych, I. (2025). The Nature of NLP: Analyzing Contributions in NLP Papers. In W. Che, J. Nabende, E. Shutova, & M. T. Pilehvar (Eds.), Proceedings of the 63rd Annual Meeting of the Association for Computational Linguistics (Volume 1: Long Papers) (pp. 25169–25191). Association for Computational Linguistics. https://doi.org/10.18653/v1/2025.acl-long.1224

Quan, Z., Mao, J., & Li, G. (2025). Multi-intent prediction of scientific literature based on heterogeneous graph fusion network. Scientometrics. https://doi.org/10.1007/s11192-025-05493-x

Qwen Team (Host). (2025, March). QwQ-32B: Embracing the Power of Reinforcement Learning [Broadcast]. https://qwenlm.github.io/blog/qwq-32b/

Safón, V., Docampo, D., & Cram, L. (2025). Screening articles by citation reputation. Quantitative Science Studies, 6, 405–420. https://doi.org/10.1162/qss_a_00355

Small, H. (1982). Citation Context Analysis. In M. J. Voigt & B. Dervin (Eds.), Progress in Communication Sciences (Vol. 3, pp. 287–310). Ablex Publishing.

Small, H. G. (1978). Cited Documents as Concept Symbols. Social Studies of Science, 8(3), 327–340. https://doi.org/10.1177/030631277800800305

Tahamtan, I., & Bornmann, L. (2019). What do citation counts measure? An updated review of studies on citations in scientific documents published between 2006 and 2018. Scientometrics, 121(3), 1635–1684. https://doi.org/10.1007/s11192-019-03243-4

Teplitskiy, M., Duede, E., Menietti, M., & Lakhani, K. R. (2022). How status of research papers affects the way they are read and cited. Research Policy, 51(4), 104484. https://doi.org/10.1016/j.respol.2022.104484

Teufel, S., & Moens, M. (2002). Summarizing Scientific Articles: Experiments with Relevance and Rhetorical Status. Computational Linguistics, 28(4), 409–445. https://doi.org/10.1162/089120102762671936

Teufel, S., Siddharthan, A., & Tidhar, D. (2006). Automatic classification of citation function. In D. Jurafsky & E. Gaussier (Eds.), Proceedings of the 2006 Conference on Empirical Methods in Natural Language Processing (pp. 103–110). Association for Computational Linguistics. https://aclanthology.org/W06-1613

Valenzuela, M., Ha, V., & Etzioni, O. (2015). Identifying Meaningful Citations.

Vital, A., Silva, F. N., Oliveira, O. N., & Amancio, D. R. (2025). Predicting citation impact of research papers using GPT and other text embeddings. Physica A: Statistical Mechanics and Its Applications, 674, 130789. https://doi.org/10.1016/j.physa.2025.130789

Wilkinson, M. D., Dumontier, M., Aalbersberg, Ij. J., Appleton, G., Axton, M., Baak, A., Blomberg, N., Boiten, J.-W., da Silva Santos, L. B., Bourne, P. E., Bouwman, J., Brookes, A. J., Clark, T., Crosas, M.,

Dillo, I., Dumon, O., Edmunds, S., Evelo, C. T., Finkers, R., … Mons, B. (2016). The FAIR Guiding Principles for scientific data management and stewardship. Scientific Data, 3(1), 160018. https://doi.org/10.1038/sdata.2016.18

Zhang, Y., Zhao, R., Wang, Y., Chen, H., Mahmood, A., Zaib, M., Zhang, W. E., & Sheng, Q. Z. (2022). Towards employing native information in citation function classification. Scientometrics, 127(11), 6557–6577. https://doi.org/10.1007/s11192-021-04242-0

---

[1] https://aclanthology.org/

[2] https://github.com/kermitt2/grobid

[3] https://aclweb.org/aclwiki/Best_paper_awards